# Noise-enhanced temporal boundary states in non-Hermitian systems

Jielong Zhang[1,2], Bozheng Xue[2], Xianfeng Man[2,3], Baizhan Xia[1,2*]

1 State Key Laboratory of Advanced Design and Manufacturing for Vehicle Body, Hunan University, Changsha, Hunan, People's Republic of China, 410082;

2 College of Mechanical and Vehicle Engineering, Hunan University, Changsha, Hunan, People's Republic of China, 410082

3 College of Mechanical and Electrical Engineering, Changsha University, Changsha, Hunan, People's Republic of China, 410022

*Email: xiabz2013@hnu.edu.cn

Time-periodic modulation introduces a synthetic degree of freedom to manipulate topological phases. This synthetic dimension can trigger a phase transition that localizes a boundary state at the temporal interface. Noise is widely deemed a fundamental threat to topological protection, universally anticipated to weaken or even destroy topological states. Here, we introduce periodically repeated temporal noise and fully random temporal noise into a 2D periodically driven non-Hermitian system. Paradoxically, under ensemble averaging, such temporal noise drives an exponential enhancement of the response intensity at the temporal interface. An averaged superoperator analysis shows that the noiseless band structure is preserved under both types of noise. Yet the noise increases the growth rate of growing modes while suppressing the decay rate of decaying ones. Finally, we experimentally realize the noise-enhanced temporal boundary state in a robotic metamaterial network. This finding establishes temporal noise as a constructive ingredient, enabling the unambiguous emergence of topological states and conferring exceptional robustness upon topological devices.

**Introduction.** Topological phases represent a major frontier in condensed-matter physics due to their robust transport properties immune to disorder and local defects. Through artificial metamaterials, topological states have successfully migrated beyond electronic systems[1–6] into classical photonic[7,8], electromagnetic[9], acoustic[10,11], electrical[12–15] and mechanical wave platforms[16–18]. Beyond spatial engineering, time−periodic modulation introduces an additional degree of freedom for manipulating topological phases, such as Floquet topological insulators[19–21], time crystals[22–29] and topological space–time crystals[30–34]. As demonstrated in photonic time crystals, temporal modulation opens a nontrivial momentum gap in the quasi-energy spectrum, whereas an abrupt temporal phase transition creates a temporal interface that hosts time-refraction and time-reflection phenomena[35]. This sudden transition between topologically disparate phases can spontaneously host a protected boundary state within the momentum gap. Spurred by the fact that non-Hermitian topologies naturally enrich complex energy bands and exceptional momentum-gap features, topological temporal boundary states have recently been realized in non-Hermitian optical[36], acoustic[37] and mechanical systems[38].

Temporal noise represents stochastic fluctuations inherent to practical time-periodic modulations. It acts as unavoidable perturbation in Floquet systems and breaks time-translation symmetry[39–42]. Recently, the Floquet–Lindblad formalism has revealed noise-induced decay of topological states[40]. It was shown that topological responses can be robust against noise over finite evolution times[41], and that chiral symmetry can further improve the stability of noisy systems[42]. To date, research has predominantly focused on mitigating the destructive influence of noise. A more counterintuitive question remains unexplored: can temporal noise, at the statistical level, provide physical enhancement of temporal boundary state in non-Hermitian Floquet systems?

In this work, we construct a 2D periodically driven non-Hermitian system. By imposing time-periodic modulation on damping terms, we clearly observe temporal boundary states in the noiseless system. On this basis, we introduce two representative types of temporal noise: periodically repeated temporal noise (PRTN) and fully random temporal noise (FRTN). Strong noise destabilizes the instantaneous response; however, the averaged response unveils a counterintuitive phenomenon: the temporal boundary states persist, and the intensity on the temporal interface increases exponentially with noise strength. These findings not only offer new insights into the statistical dynamics of non-Hermitian Floquet topological systems, but also establish temporal noise as a constructive mechanism for topological amplification, enabling the enhancement of otherwise elusive interface states.

**Tight-Binding Model.** We consider a 2D tight-binding model with staggered gain and loss, subjected to a time-periodic square-wave modulation. As illustrated in Fig. 1a, the model features uniform intra- and inter-cell hopping amplitudes along both the $x$ and $y$ directions, except for a sign reversal of the hopping along the $y$ direction in even-numbered columns. Meanwhile, the on-site gain and loss terms are arranged alternately in a counterclockwise sequence within a four-site unit cell. The Hamiltonian of the model can be written as:

$$\boldsymbol{H}(t)=\begin{bmatrix} i(\Upsilon_s+\Upsilon_d(t)) & w(1+e^{-iq_x}) & 0 & w(1+e^{-iq_y}) \\ w(1+e^{iq_x}) & -i(\Upsilon_s+\Upsilon_d(t)) & -w(1+e^{-iq_y}) & 0 \\ 0 & -w(1+e^{iq_y}) & i(\Upsilon_s+\Upsilon_d(t)) & w(1+e^{iq_x}) \\ w(1+e^{iq_y}) & 0 & w(1+e^{-iq_x}) & -i(\Upsilon_s+\Upsilon_d(t)) \end{bmatrix} \quad (1)$$

Here, $\pm\Upsilon_s$ denotes the static gain-loss term and $\Upsilon_d=\alpha\mathrm{sgn}(\cos(\Omega t))$ represents the dynamically modulated gain-loss term. The parameter $\alpha$ is a constant that characterizes the driving strength, $\Omega=2\pi/T$ is the modulation frequency, $T$ is the modulation period, and $w$ denotes the hopping strength. The Hamiltonian $\boldsymbol{H}(t)$ satisfies time-translation symmetry, namely $\boldsymbol{H}(t)=\boldsymbol{H}(t+T)$. We set $w=1$, $\Upsilon_s=0.3$, and $\Omega=3.6$. The quasi-energy spectrum is obtained from the Floquet effective Hamiltonian $\boldsymbol{H}_{eff}=\frac{1}{i2\pi T}\ln\boldsymbol{U}(T)$. $\boldsymbol{U}(T)=e^{-i2\pi\mathcal{T}\int_0^T \boldsymbol{H}(t)dt}$ is the time-evolution operator over a modulation period, where $\mathcal{T}$ denotes the time-ordering operator. As shown in Fig. 1b and d, when $\alpha=\mp0.2$, four annular momentum gaps emerge within the quasi-energy spectrum. In these gaps, the imaginary part of the quasi-energy $\varepsilon$ is nonzero[37]. By contrast, as shown in Fig. 1c, in the absence of time-periodic modulation ($\alpha=0$), the momentum gaps collapse.

We can obtain the low energy effective Hamiltonian $H_{eff}$ near the gap (See Supplementary Note 1)

$$\boldsymbol{H}_{eff}=\frac{\Omega}{2}\boldsymbol{I}_2+v_D\boldsymbol{q}_\perp\boldsymbol{\tau}_z-im_D\boldsymbol{\tau}_y, \quad (2)$$

where $\boldsymbol{I}_2$ denotes the $2\times2$ identity matrix, $v_D=\frac{w^2}{\varepsilon_0}\sqrt{\sin^2 q_0^x+\sin^2 q_0^y}$ denotes the effective Dirac velocity, $\varepsilon_0$ denotes the quasi-energy of $\pi$-gap, $q_0^x$ ($q_0^y$) denotes the momentum component along $x$ ($y$) direction in $\pi$-gap, $\boldsymbol{q}_\perp$ denotes the normal momentum offset, $\boldsymbol{\tau}_y$ ($\boldsymbol{\tau}_z$) denotes the Pauli matrix, and $m_D=\frac{2\alpha}{\pi}\sqrt{1+\frac{4{\Upsilon_s}^2}{\Omega^2}}$ denotes the effective Dirac mass. As the system transitions from the $\alpha<0$ to the $\alpha>0$ regime in Fig. 1e, the effective Dirac mass undergoes a sign reversal from negative to positive. Meanwhile, at the Floquet momentum gap, reversing the sign of $\alpha$ inverts the growing and decaying states of eigenvectors $\boldsymbol{\psi}_1$, $\boldsymbol{\psi}_2$, $\boldsymbol{\psi}_3$, and $\boldsymbol{\psi}_4$. As illustrated in Fig. 1f, the imaginary parts of quasienergies undergo a reversal across $\alpha=0$, demonstrating a mutual exchange between the amplifying and attenuating modes. Consequently, these two regimes ($\alpha<0$ and $\alpha>0$) constitute temporally adjacent domains with distinct topological phases[37,38,43–46], naturally fostering a temporal boundary state (See Supplementary Note 1).

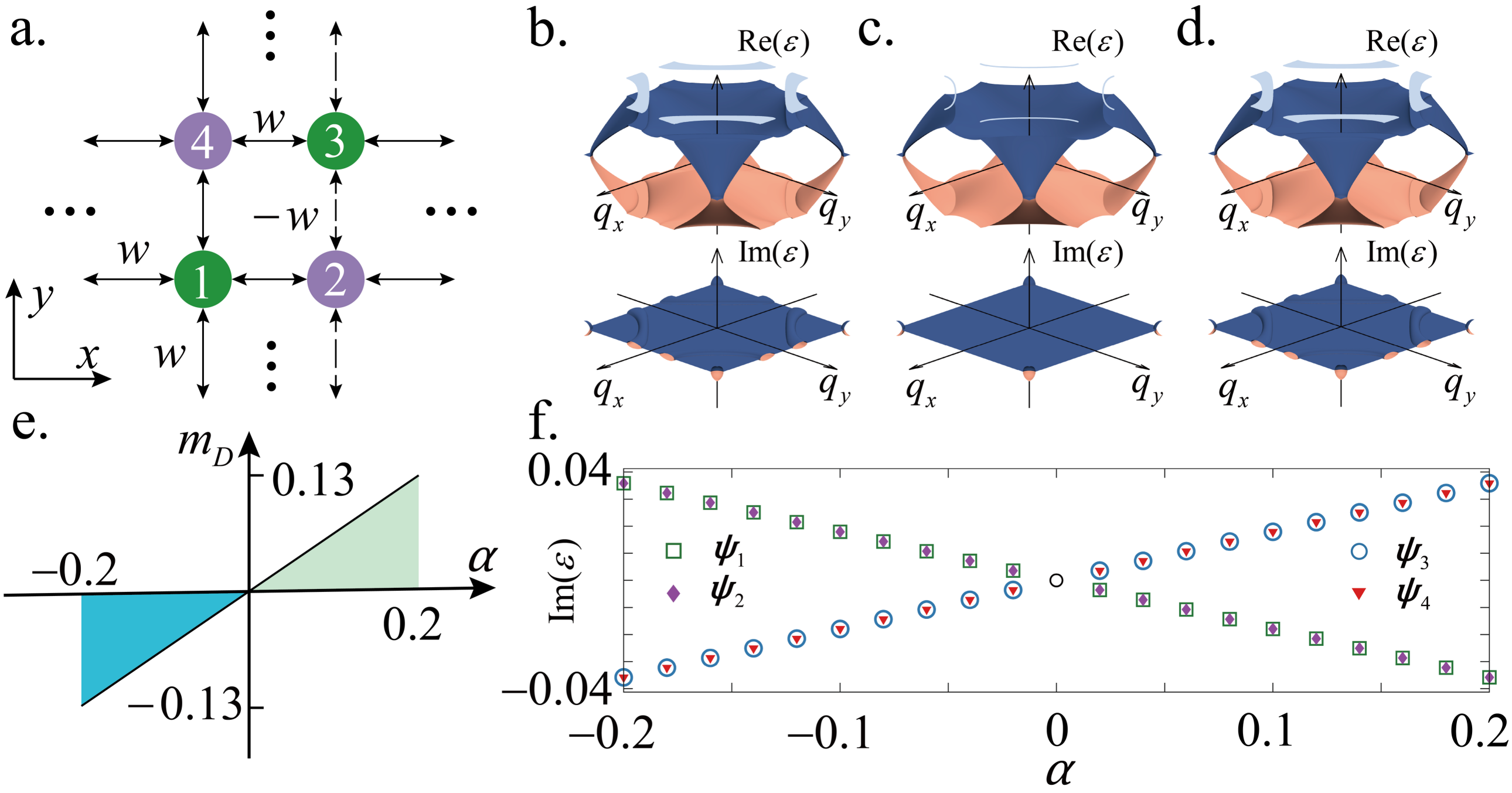

**Fig. 1 | Time-periodically modulated 2D tight-binding model. a,** Configuration of unit cell with uniform intra- and inter-cell couplings. Dashed lines denote sign-reversed couplings along the *y*-direction in even-numbered columns. Green and purple sites represent gain and loss, respectively. **b–d,** Real and imaginary quasi-energy spectra under time-periodic modulation. Brown and dark blue regions denote negative and positive quasi-energies, respectively. When $\alpha = -0.2$(b) and $\alpha = 0.2$(d), nontrivial momentum gaps open, within which the imaginary parts of the quasi-energies become non-zero. **e**, Effective Dirac mass $m_D$ as a function of $\alpha$. **f,** Imaginary parts of quasi-energies within the momentum gap as a function of $\alpha$. Blue circles, red triangles, green squares, and purple diamonds denote eigenvectors $\boldsymbol{\psi}_1$, $\boldsymbol{\psi}_2$, $\boldsymbol{\psi}_3$, and $\boldsymbol{\psi}_4$, respectively.

**Multi-degree-of-freedom spring–mass network.** Analogous to the 2D tight-binding model, we construct a multi-degree-of-freedom spring–mass network[47]. As illustrated in Fig. 2a, each oscillator is grounded via a spring with equivalent stiffness $k_1$, while adjacent oscillators are coupled through interconnecting springs with equivalent stiffness $k_2$. According to Bloch's theorem, the displacement of a lattice site in a spatially translated unit cell is identical to that of its intra-cell counterpart, offset merely by a spatial phase factor. Specifically, the spatial translation of displacements along the positive and negative axes yields the Bloch phase factors $e^{iq_x}$ $(e^{iq_y})$ and $e^{-iq_x}$ $(e^{-iq_y})$, respectively, where $q_x$ and $q_y$ denote the momentum components in the $x$ and $y$ directions, respectively. To synthesize equivalent periodic boundaries[37], within an isolated unit cell, we introduce two pairs of phase-modulated couplings, explicitly parameterized as $k_2 e^{\pm i q_x}$ and $k_2 e^{\pm i q_y}$. These couplings emulate the equivalent relations between oscillators and their nearest neighbors in adjacent unit cells.

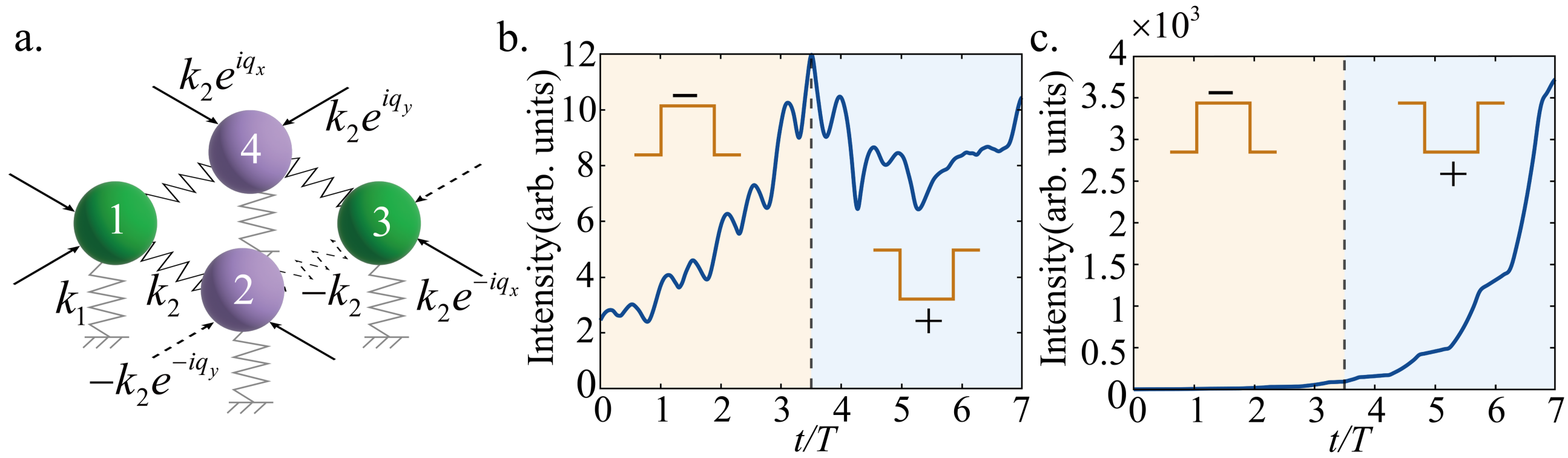

**Fig. 2 | Temporal boundary state in a multi-degree-of-freedom spring–mass network driven by time-periodic square-wave modulation. a,** Schematic of the network. Green and purple oscillators feature active gain (negative damping) and loss (positive damping), respectively, synchronized via a time-periodic square-wave. Each oscillator possesses an intrinsic equivalent stiffness $k_1$. Dashed lines indicate negative equivalent stiffness. **b,** Temporal boundary state. The periodic modulations flanking the temporal interface at $t$=3.5$T$ possess a π-phase shift. **c,** Response away from Floquet momentum gap.

The dynamical response of the multi-degree-of-freedom spring-mass network is governed by a second-order equilibrium equation. To map into a first-order canonical form compatible with the Floquet-Schrödinger formalism, we set $k_1 \gg k_2$. The network is excited by a Gaussian-envelope force $F = e^{-(t-t_p)^2/2t_\sigma^2}\cos(\omega_0 t)$, where $t_p$ and $t_\sigma$ parameterize the temporal center and duration of the envelope, respectively, and $\omega_0 = \sqrt{k_1/I}$ denotes the intrinsic resonant frequency. Under the slowly varying envelope approximation ($|\dot{\boldsymbol{A}}(t)| \ll \omega_0|\boldsymbol{A}(t)|$), the collective displacement vector can be factorized as $\boldsymbol{\Theta}(t) = \boldsymbol{A}(t)e^{i\omega_0 t}$. Here, $\boldsymbol{\Theta} = [\theta_1(t), \theta_2(t), \theta_3(t), \theta_4(t)]^T$ captures the instantaneous site displacements, and $\boldsymbol{A}(t) = [A_1(t), A_2(t), A_3(t), A_4(t)]^T$ represents the slowly varying modulation amplitudes. In this asymptotic limit, the second-order governing equation reduces to a first-order one, yielding the non-Hermitian time-evolution equation[48–50] written as (see Supplementary Note 2)

$$i\frac{d\boldsymbol{A}}{dt} = \boldsymbol{HA}, \tag{3}$$

where $\boldsymbol{H}(t) = \begin{bmatrix} \zeta & -\xi(1+e^{-iq_x}) & 0 & -\xi(1+e^{-iq_y}) \\ -\xi(1+e^{iq_x}) & -\zeta & -\xi(1+e^{-iq_y}) & 0 \\ 0 & -\xi(1+e^{iq_x}) & \zeta & -\frac{k_2}{2k_1}\omega_0(1+e^{iq_y}) \\ -\xi\omega_0(1+e^{iq_y}) & 0 & -\xi(1+e^{-iq_x}) & -\zeta \end{bmatrix}$

Here, $\zeta = \omega_0 - i\frac{\gamma(t)}{2I} + \frac{k_2}{2k_1}\omega_0, \xi = \frac{k_2}{2k_1}\omega_0$ . we set $I = 1\text{kg}\cdot\text{m}^2$ , $k_1 = 1.5$ N/rad , $k_2 = 0.0556$ N/rad, $\gamma(t) = 0.0131 + 0.0088\text{sgn}(\cos(\Omega t))$, and $\Omega = 0.0788$ rad/s. We consider $q_x = 0.35\pi$ and $q_y = 0.8\pi$, where the quasi-energy is within the momentum gap and has a nonzero imaginary part. A Gaussian-envelope pulse $F$ with a duration of $1s$ is then launched at the lower-left oscillator. As shown in Fig. 2b, within the temporal window spanning from $t = 0s$ to $t = 3.5T$, the envelope intensity of four oscillators in a unit cell exhibits a clear exponential growth. After $t = 3.5T$, we suddenly invert the modulation phase by reversing the sign of the time-periodic damping component, namely $\gamma(t) = 0.0131 - 0.0088\text{sgn}(\cos(\Omega t))$. The dynamical response undergoes a sharp transition from a growing mode to a decaying mode, manifesting as an exponential attenuation of the wave amplitude. The envelope intensity precisely peaks at the temporal boundary $t$=3.5$T$, establishing a well-defined temporal boundary state. Notably, immediately following this peak, the response first undergoes a transient exponential decay and then gives way to a re-amplification. This behavior stems from the simultaneous excitation of both the growing and decaying modes of the non-Hermitian system; and the contribution from the growing modes eventually dominates over that from the decaying modes[36]. When the momentum ($q_x = \pi$ and $q_y = \pi$) resides outside the momentum gap, as illustrated in Fig. 2c, the dynamical response grows exponentially over time without forming a localized peak at the temporal interface, highlighting a contrast to the interface-localized response observed within the momentum gap.

**Noise-enhanced temporal boundary states in multi-degree-of-freedom spring–mass networks**

We first evaluate the spring–mass network subjected to PRTN. As illustrated in Fig. 3a, each driving period $T$ is partitioned into 40 subintervals. Within the $j$-th subinterval, we introduce a stochastic variable $r_j$ , drawn from a uniform distribution $r_j \in [-W_T, W_T]$ . Accordingly, the instantaneous damping profile is expressed as $\gamma_j(t) = 0.0131 \pm 0.0088\text{sgn}(\cos(\Omega t))(1 + r_j)$ , where the sign of the time-varying damping component remains strictly inverted across the temporal interface. Within a single realization, the noise profile is constrained by time-translation symmetry, such that the random noise pattern repeats identically across consecutive periods, satisfying $\gamma_j(t+T)=\gamma_j(t)$. We investigate the dynamic response $p(t,W_T)$ across distinct noise strengths $W_T\in\{1,2,\dots,8\}$. For each noise strength, 1000 independent stochastic realizations are generated. Although individual responses exhibit strong random fluctuations heavily perturbed by the noise (Supplementary Note 5), the averaged intensity of response $f(t,W_T)=E[p(t,W_T)]$, reveals a counterintuitive phenomenon: the temporally localized state is robustly amplified by the temporal noise (Fig. 3b). As shown in Fig. 3c, the expectation of the response intensity ($\mu(\mathrm{W_T})=f(3.5T,W_T)$) at the temporal interface scales exponentially with the noise strength, obeying $\mu(W_T)\sim 0.579e^{0.462W_T}+11.403$. Concurrently, the standard deviation ($\sigma(W_T)=\sqrt{\mathrm{var}[p(3.5T,W_T]}$) at the temporal interface mimics this exponential increasing, namely $\sigma(W_T)\sim 0.019e^{0.953W_T}-0.019$.

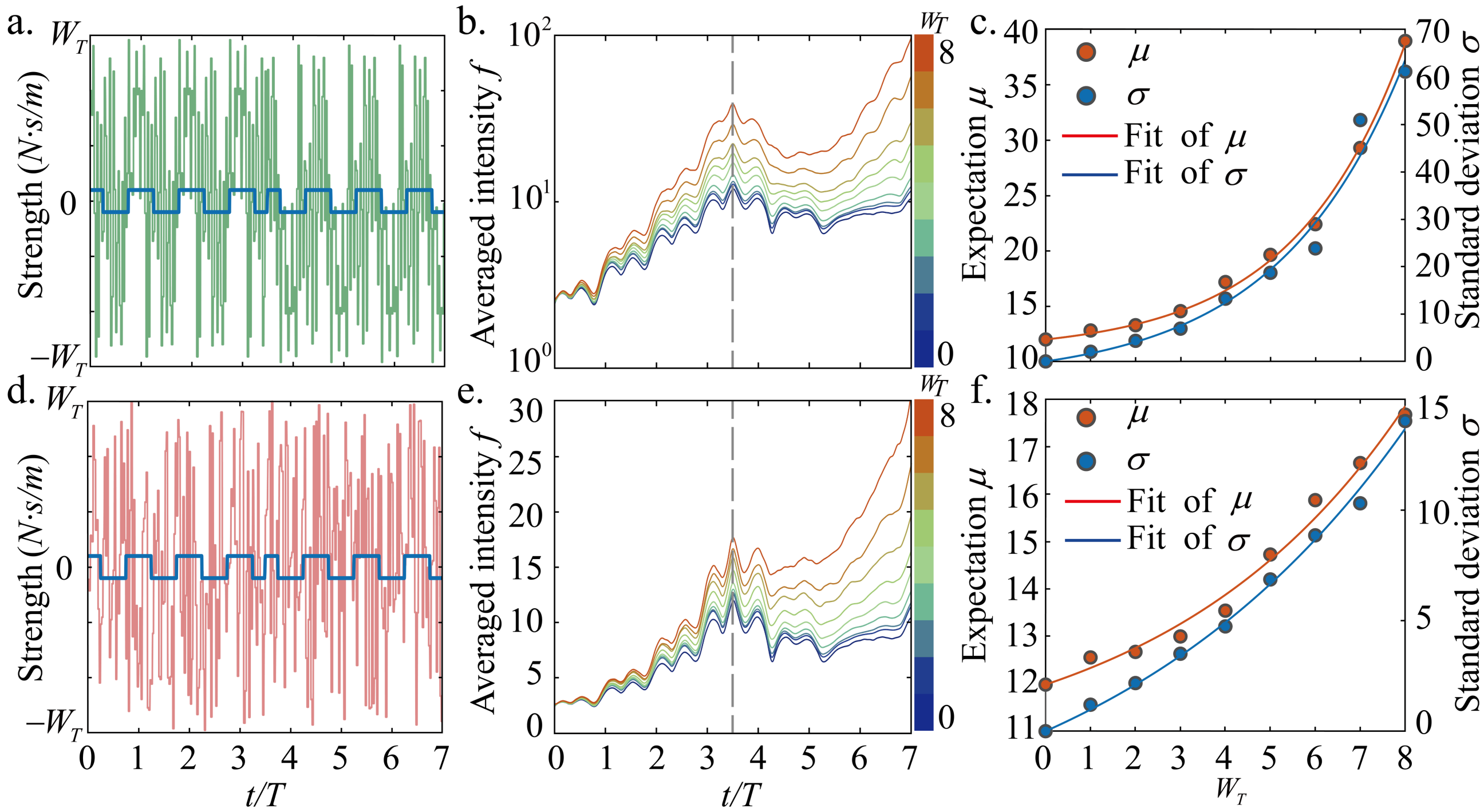

**Fig. 3 | Noise-enhanced temporal boundary states under PRTN and FRTN. a**, PRTN. Random noise $r_j\in[-W_T,W_T]$ are frozen within one period but repeat identically across consecutive periods. **b**, Averaged responses for PRTN with different strengths $W_T$. **c**, Expectations ($\mu$) and standard deviations ($\sigma$) of the response intensity at the temporal interface as a function of $W_T$. **d**, FRTN, which is stochastically resampled at every modulation period. **e**, Averaged responses for FRTN with different strengths $W_T$. **f**, Expectations ($\mu$) and standard deviations ($\sigma$) of the response intensity at the temporal interface as a function of $W_T$. In **b** and **e**, the color gradient from blue to orange

tracks increasing $W_T$ (from 0 to 8), with the grey dashed line marking the temporal interface. In **c** and **f**, symbols and solid lines represent simulated results and corresponding exponential fits, respectively.

Next, we replace the PRTN with FRTN. As illustrated in Fig. 3d, the noise within each modulation period is sampled independently, while following the same uniform distribution. This system explicitly breaks time-translation symmetry, such that $\gamma_j(t+T) \neq \gamma_j(t)$. We evaluate the dynamical responses for varying noise strengths $W_T \in \{1, 2, \ldots, 8\}$. Mirroring the behavior observed under PRTN, the single response is highly randomized, with their fluctuation bounds expanding alongside the noise strength (Supplementary Note 5). Nevertheless, the averaged response $f(t, W_T)$ remarkably preserves the temporal boundary state (Fig. 3e). The expectation of the response intensity $\mu(W_T)$ at the temporal interface grows exponentially with noise strength (Fig. 3f), obeying $\mu(W_T) \sim 1.47e^{0.206W_T} + 10.512$. Concurrently, the standard deviation of the peak intensity ($\sigma(W_T)$) also exhibits an exponential growth, parameterized as $\sigma(W_T) \sim 5.229e^{0.163W_T} - 5.229$.

The noise-enhanced temporal boundary state under PRTN and FRTN can be analytically captured by the averaged superoperator[41]. We first define an outer-product matrix, $\boldsymbol{\Pi}(t) = |\boldsymbol{A}(t)\rangle\langle\boldsymbol{A}(t)|$. The trace $Tr[\boldsymbol{\Pi}(t)] = \sum_{n=1}^{4}|\boldsymbol{A}_n(t)|^2$ explicitly yields the response intensity confined within the unit cell. By vectorizing $\boldsymbol{\Pi}(t)$ into a vector $\mathrm{vec}[\boldsymbol{\Pi}(t)]$, its time evolution can be formulated as $i\frac{d}{dt}\mathrm{vec}[\boldsymbol{\Pi}(t)] = \boldsymbol{\mathcal{H}}\mathrm{vec}[\boldsymbol{\Pi}(t)]$. Here, the superoperator is defined as $\boldsymbol{\mathcal{H}} = \boldsymbol{I}_4 \otimes \boldsymbol{H} - \boldsymbol{H}^T \otimes \boldsymbol{I}_4$, where $\boldsymbol{I}_4$ denotes the 4×4 identity matrix and $\otimes$ represents the Kronecker product. The propagator for this superoperator over one modulation cycle is given by $\boldsymbol{\mathcal{U}} = e^{-i\boldsymbol{\mathcal{H}}T}$. We then introduce the averaged superoperator, $\boldsymbol{\mathcal{F}} = E(\boldsymbol{\mathcal{U}})$, to characterize the statistical dynamics of the random system. A comprehensive derivation of this averaged superoperator is detailed in Supplementary Note 4, where $\boldsymbol{\mathcal{F}}$ can be written as

$$\boldsymbol{\mathcal{F}} = \int_{-W_T}^{W_T} \boldsymbol{\mathcal{U}}_{40}\, d\gamma \ldots \int_{-W_T}^{W_T} \boldsymbol{\mathcal{U}}_j\, d\gamma \ldots \int_{-W_T}^{W_T} \boldsymbol{\mathcal{U}}_1\, d\gamma, \tag{4}$$

where $\boldsymbol{U}_j = e^{-i\boldsymbol{\mathcal{H}}_j dt}$ denotes the propagator corresponding to the *j*-th noisy subinterval, *j*=1, 2, …, 40. Using the averaged superoperator $\boldsymbol{\mathcal{F}}$, the statistically averaged evolution of $\boldsymbol{\Pi}(t)$ over a full modulation period *T* can be expressed as $\overline{vec[\boldsymbol{\Pi}(T)]} = \boldsymbol{\mathcal{F}}\overline{vec[\boldsymbol{\Pi}(0)]}$. Therefore, for a noise strength $W_T$, the averaged response $f_{tb}(T)$ can be written as:

$$f_{tb}(T) = vec(\boldsymbol{I}_4)^T\overline{vec[\boldsymbol{\Pi}(t)]} = vec(\boldsymbol{I}_4)^T \sum_{n=1}^{16} \chi_n |\boldsymbol{\beta}_n^R\rangle\langle\boldsymbol{\beta}_n^L|\, vec[\boldsymbol{\Pi}(0)]. \tag{5}$$

Here, $\chi_n$ represents the eigenvalue of the averaged superoperator $\boldsymbol{\mathcal{F}}$ with noise strength $W_T$, and $\langle\boldsymbol{\beta}_n^L|$ and $|\boldsymbol{\beta}_n^R\rangle$ denote the left and right eigenvectors, respectively, and $\langle\boldsymbol{\beta}_n^L|\boldsymbol{\beta}_n^R\rangle = 1$.

To elucidate the mechanism underlying the noise-enhanced temporal boundary state, we evaluate the eigenvalues of $\boldsymbol{\mathcal{F}}$ with $q_x = 0.35\pi$ and $q_y = 0.8\pi$. As shown in Fig. 4a, the eigenvalues are purely real. With increasing noise strength, the overall configuration of the eigenvalue spectrum is preserved; however, as depicted in Fig. 4b, all eigenvalues shift toward larger positive values. Specifically, the eigenvalue corresponding to the amplifying mode departs from the unit circle, scaling exponentially with noise strength. Concurrently, the eigenvalue associated with the decaying mode initially exhibits a similar exponential growth; however, it ceases upon intersecting the unit circle, becoming strictly pinned at the boundary (Supplementary Note 4). The regression lines for the eigenvalues inside (blue points), on (red points) and outside (green points) the unit circle are $\chi_1 = 0.009e^{0.254W_T} + 0.629$, $\chi_2 = 0.008e^{0.277W_T} + 0.992$, and $\chi_3 = 0.016e^{0.307W_T} + 1.549$, respectively. According to Eq. (5), the evolution of the eigenvalue accounts for the exponential growth of the peak intensity expectation $\mu(W_T)$ with the noise strength.

Moreover, in each stochastic realization of PRTN, a random sequence is generated within a single period and then repeated identically across all subsequent periods. By contrast, within the FRTN system, the noise in different periods is sampled independently. Consequently, the multi-period averaged superoperators, denoted as $\boldsymbol{\mathcal{F}}_1^{(\mathrm{N})}$ for PRTN and $\boldsymbol{\mathcal{F}}_2^{(\mathrm{N})}$ for FRTN, are not equivalent (Supplementary Note 4), where *N* denotes the period number. We investigate the eigenvalues of the averaged superoperators $\boldsymbol{\mathcal{F}}_1^{(\mathrm{N})}$ and $\boldsymbol{\mathcal{F}}_2^{(\mathrm{N})}$ for *N*=4. As the noise strength $W_T$ increases, the eigenvalue ($\chi_N$) spectra for both $\boldsymbol{\mathcal{F}}_1^{(\mathrm{N})}$ and $\boldsymbol{\mathcal{F}}_2^{(\mathrm{N})}$ retain the foundational organization of the noiseless system (Fig. 4c); however, the eigenvalues inside, on, and outside the unit circle exhibit a more pronounced outward shift along the positive real axis compared to the single-period eigenvalues of $\boldsymbol{\mathcal{F}}$. Their deviations from the noiseless benchmarks scale exponentially with the noise strength. As shown in Figs. 4d–f, PRTN drives the evolution of eigenvalues significantly faster than its FRTN counterpart. These results elucidate why the averaged response maintains an exponential growth under FRTN, albeit experiencing a mitigated enhancement compared to the robust amplification observed in the PRTN configuration.

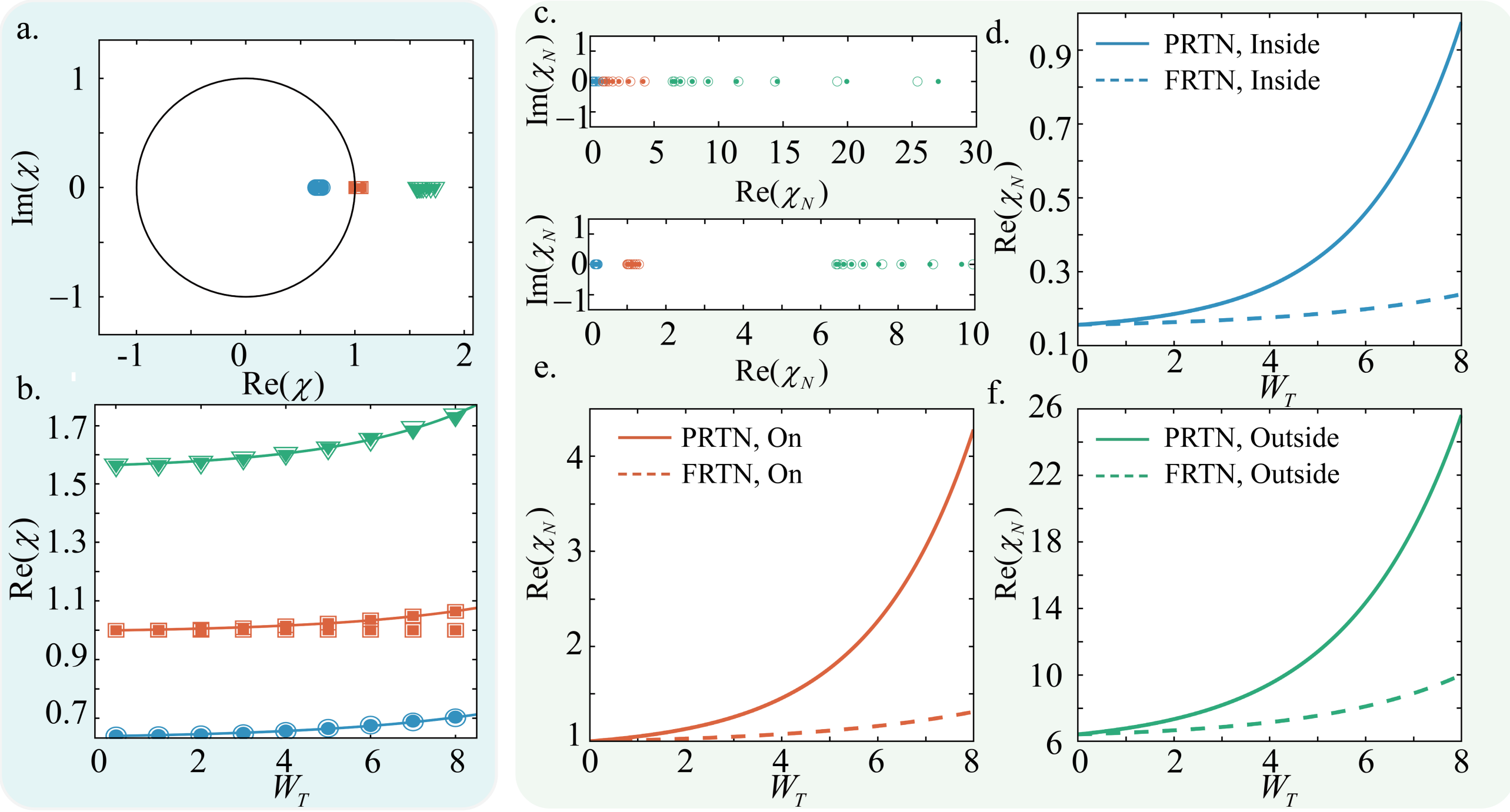

**Fig. 4 | Eigenvalues of the averaged superoperator. a**, Eigenvalue evolution of the single-period averaged superoperator $\boldsymbol{\mathcal{F}}$ under temporal noise. Blue circles, brown squares, and green triangles denote eigenvalues inside, on, and outside the unit circle, respectively. Solid and hollow markers represent a conjugate pair of networks carrying opposite time-varying damping profiles. **b**, Exponential scaling of eigenvalues (inside, on, and outside the unit circle) as a function of $W_T$. **c**, Eigenvalue spectra of the multi-period averaged superoperators for PRTN (upper panel) and FRTN (lower panel) with $N$=4. **d–f**, Exponential scaling of eigenvalues located inside (d), on (e), and outside (f) the unit circle, respectively. Solid and dashed lines correspond to PRTN and FRTN, respectively.

### Experimental realization of noise-enhanced temporal boundary states.

The simulated analysis presented above establishes that temporal noise statistically amplifies temporal boundary states. To experimentally validate this counter-intuitive phenomenon, we design a robotic metamaterial network, as shown in Fig. 5a. Its effective moment of inertia, stiffness, and both static and time-varying damping are programmed via a unified control framework. Four independently controlled robotic metamaterials serve as the constituent oscillators modeled in Fig. 2a, enabling the programmable synthesis of active generalized forces, including virtual non-reciprocal springs and tailored dampings. The prescribed coupling interactions among distinct robotic metamaterials are mediated via a communication network and built-in dynamical balance equations (see Supplementary Note 3).

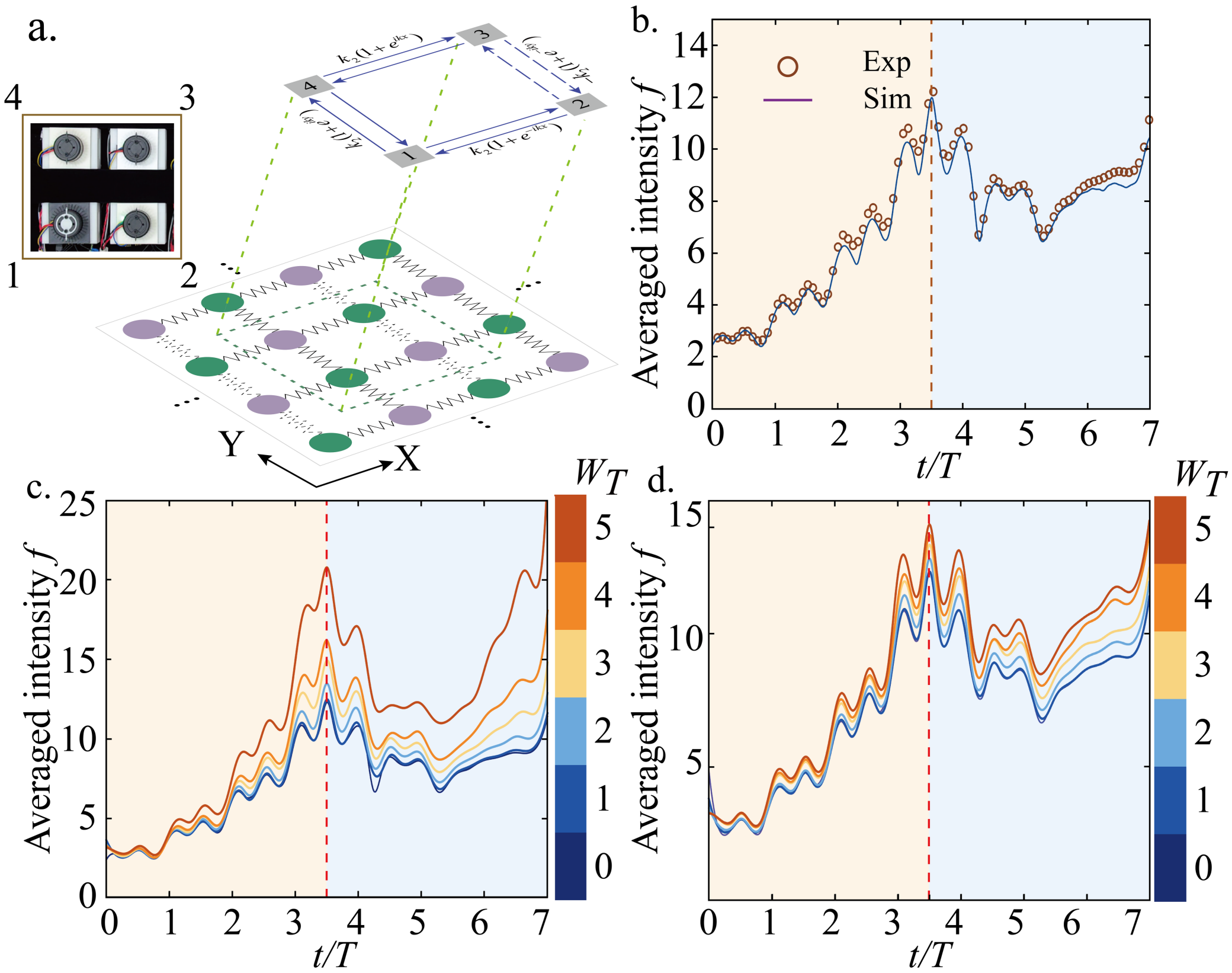

**Fig. 5 Experimental realization of noise-enhanced temporal boundary states in a robotic metamaterial network. a,** Schematic and photograph of a robotic metamaterial network. Blue arrows indicate the non-reciprocal couplings mediated between robotic metamaterials, with dashed lines denoting negative couplings. **b,** Measured response of robotic metamaterials. Red circles denote the measured response intensity, while the blue solid line denotes the simulated one. **c** and **d,** Averaged intensities of robotic metamaterials under PRTN (c) and FRTN (d) with noise strengths $W_T \in \{1, 2, \dots, 5\}$.

As demonstrated in Fig. 5b, upon launching a Gaussian-envelope excitation to the lower-left robotic metamaterial, the measured response develops a pronounced peak at the temporal interface. The domains flanking this interface exhibit clear amplifying and decaying modes, respectively, confirming the formation of the temporal boundary state, which is in excellent agreement with the simulated results. We then actively superimpose either PRTN or FRTN onto the time-periodic square-wave, with noise strengths $W_T \in \{1, 2, \dots, 5\}$ (detailed in supplementary Note 3). Although a single stochastic experiment indicates that noise may weaken the temporal boundary state (see Extended Data Figs. 1 and 2), the averaged response curves displayed in Figs. 5c and d unambiguously demonstrate that the temporal boundary states are robustly amplified by the temporal noise. Specifically, the expectation at the temporal interface scales from 12.3 in the noiseless baseline to 20.78 ($W_T = 5$) under PRTN, and to 14.1 ($W_T = 5$) under FRTN. Consistent with our superoperator analysis (Fig. 4c), FRTN exerts a mitigated amplification effect on the Floquet-modulated lattice compared to its PRTN counterpart.

**Conclusion.**

In summary, we have demonstrated a temporally modulated robotic metamaterial network and uncovered a counterintuitive enhancement effect driven by temporal noise on temporal boundary states. In sharp contrast to the conventional results where noise inevitably degrades topological protection, we demonstrate that temporal noise can actively induce an exponential scaling of the localized interface intensity under ensemble-averaged dynamics. The averaged superoperator explicitly elucidates the physical mechanism underlying this noise-enhanced manifestation. These findings profoundly reshape the current understanding of the interplay between noise and non-Hermitian topology in Floquet systems. As our theoretical formalism is rooted in foundational Schrödinger dynamics and Floquet engineering, the noise-amplified temporal boundary states are highly versatile, directly transferable from active robotic metamaterials to photonic, acoustic, and quantum architectures. Crucially, leveraging such noise-enhancing effectively transforms noise from a detrimental nuisance into a functional asset, offering a powerful alternative for signal amplification, and thereby establishing a new design blueprint for high-performance, noise-resilient topological devices.

## Methods

**Wave-function overlap matching**

To track the four eigenvectors across the parameter space, we employ a state-tracking method based on wavefunction overlap matching (detailed Supplementary Note 1). By maximizing the trace of the overlap matrix, we establish a one-to-one correspondence between the two sets of eigenvectors, thereby identifying the quasienergies associated with the same eigenstates across the two distinct regimes ($\alpha < 0$ and $\alpha > 0$). This identifies the topological transition by directly tracking the evolution of physical eigenstate branches. In the parameter regime, $|\alpha|/w \ll 1$ and $\Omega/w \gg 1$, the coupling between Floquet sectors satisfies the high-frequency and weak-driving condition. The off-diagonal coupling elements attenuate rapidly as the Floquet-sector index difference $\Delta m$ increases. The interaction between neighboring Floquet sectors remains non-negligible, while coupling across higher-order sectors asymptotically vanishes. Therefore, truncating the Floquet extended Hamiltonian to the lowest relevant neighboring sectors is sufficient to accurately describe the $\pi$-gap modes and their topological properties. We select a momentum point $\boldsymbol{q}_0$ on the Dirac nodal line and evaluate the Floquet extended Hamiltonians by sweeping the parameter $\alpha$ from $-0.2$ to $0.2$ in discrete increments of $0.02$. Diagonalizing these Hamiltonians, we obtain the corresponding quasienergy spectra and eigenvectors $|\boldsymbol{\psi}_n\rangle$, where *n* represents the *n*-th step. For the four $\pi$-gap modes under the *n*-step and the *n*+1-step, we construct the overlap matrix $\boldsymbol{M} = |\langle\boldsymbol{\psi}_n^L|\boldsymbol{\psi}_{n+1}^R\rangle\langle\boldsymbol{\psi}_n^L|\boldsymbol{\psi}_{n+1}^R\rangle|$, where $\langle\boldsymbol{\psi}_n^L|$ and $|\boldsymbol{\psi}_{n+1}^R\rangle$ represent the left eigenvector of the *n*-step and the right eigenvector of the *n*+1-step, respectively. The optimal pairing between the two sets of eigenvectors is determined by maximizing the trace of the matched overlap matrix. Based on this sorting scheme, we can unambiguously track each individual eigenvector and the sign variation of the imaginary quasienergies associated with parametric evolution.

**The noise generation**

The active noise generation module consists of a STM32F405RGT6 microcontroller and its peripheral supporting circuitry. We use its random-number generator to yield random seeds. These seeds are used to generate the random numbers. The random sequences remain statistically independent across different realizations while strictly obeying the identical uniform distribution over $[0, 1]$. The generated random numbers are subsequently map-transformed to conform to a physical uniform distribution over $[-W_T, W_T]$. Different noise generation strategies are used for PRTN and FRTN. For PRTN, the microcontroller synthesizes an immutable array of 40 random numbers at the experimental onset, which is then cyclically replayed at each consecutive modulation period. Conversely, for FRTN, the module dynamically yields new random numbers for all modulation periods.

**Experimental test.**

The robotic metamaterials are actuated by MS4005 brushless motors integrated with high-precision sensing and driving modules, featuring a fine angular resolution of 0.01°. To ensure deterministic real-time execution, each robotic metamaterial deploys a dual-core architecture consisting of two STM32F103C8T6 microcontroller units (MCUs). Specifically, MCU 1 is dedicated to executing the governing torque-displacement constitutive equations to produce the instantaneous displacement and velocity in real time. In addition, MCU1 also manages inter-node cooperative data exchange via a universal asynchronous receiver-transmitter (UART) protocol, continuously transmitting local kinematic states (including angular and velocity) to neighboring robotic metamaterials. Concurrently, MCU 2 captures the value of external stochastic noise from the noise generation module and routes them to MCU 1 via a dedicated parallel bus. The deterministic control loop is strictly scheduled with a temporal step of 25 ms.

**Acknowledgements:** This work was supported by the Natural Science Foundation of Hunan Province for Young Scientists Fund (Category A, including continuation funding projects) (Grant No. 2026JJ20001), the National Natural Science Foundation of China (Grant Nos. 52475254 and 12072108), and the Project of State Key Laboratory of Advanced Design and Manufacturing for Vehicle Body.

**Author contributions:** J.Z. and B.X. designed the research. J.Z., B.X., and X.M. conducted theoretical analysis. J.Z. and B.X. conducted the measurements. J.Z. and B.X. analyzed the data. J.Z., B.X., X.M., and B.X. wrote the paper. B.X. supervised the project.

**Data availability:** The data that support the findings of this article are not publicly available. The data are available from the authors upon reasonable request.

**Competing interests:** The authors declare no competing interests.

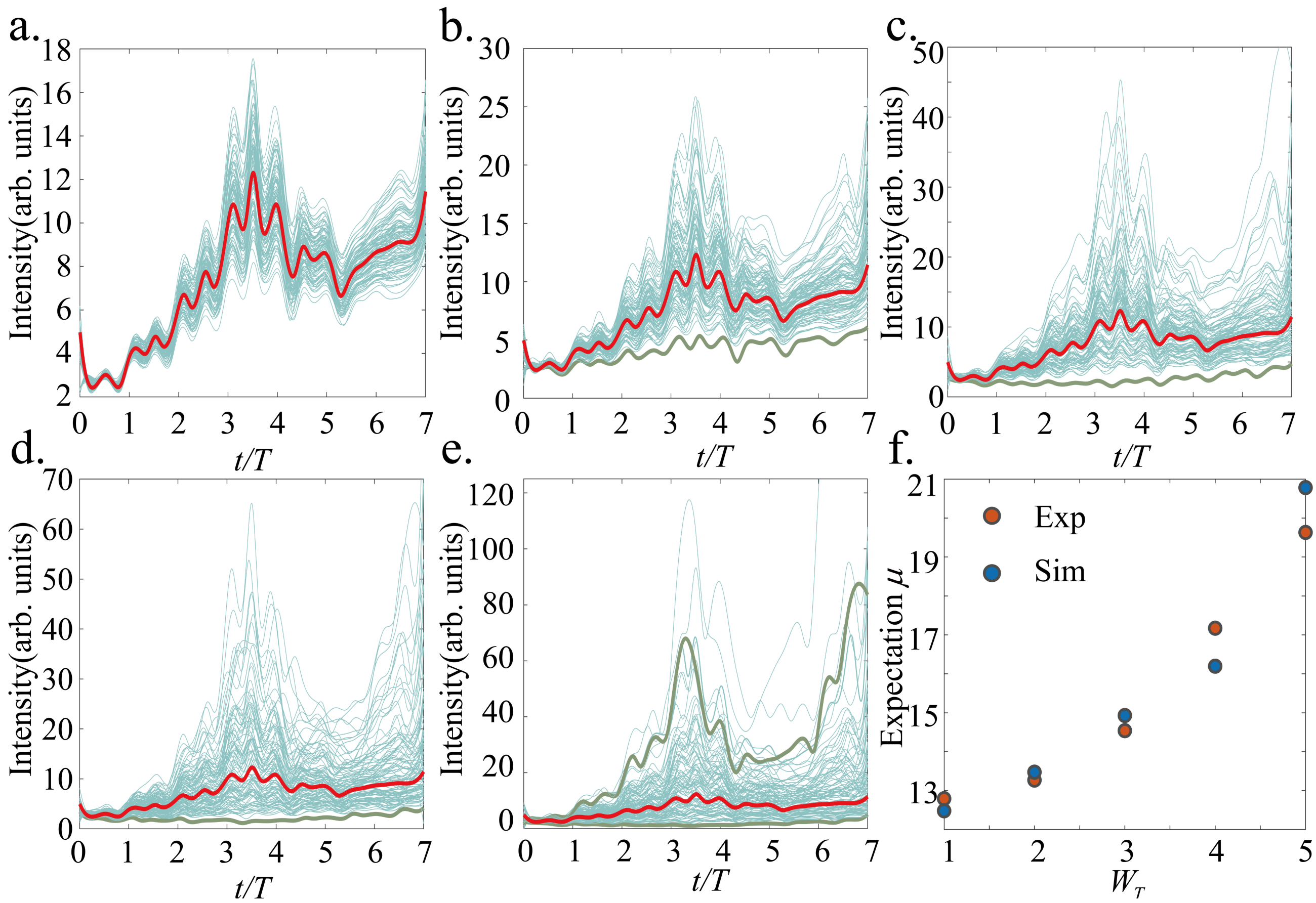

Extended Data Figure 1. Experimental dynamical responses under PRTN. **a**-**e**, Dynamic responses from individual experimental realizations corresponding to noise strengths $W_T = 1$ (**a**), $W_T = 2$ (**b**), $W_T = 3$ (**c**), $W_T = 4$ (**d**), $W_T = 5$ (**e**). The red line designates the deterministic response of the noiseless system. Driven by the temporal noise, the responses fluctuate randomly around the noiseless baseline. **f**, Comparison of the intensity expectations between the experimental measurements and numerical simulations as a function of $W_T$.

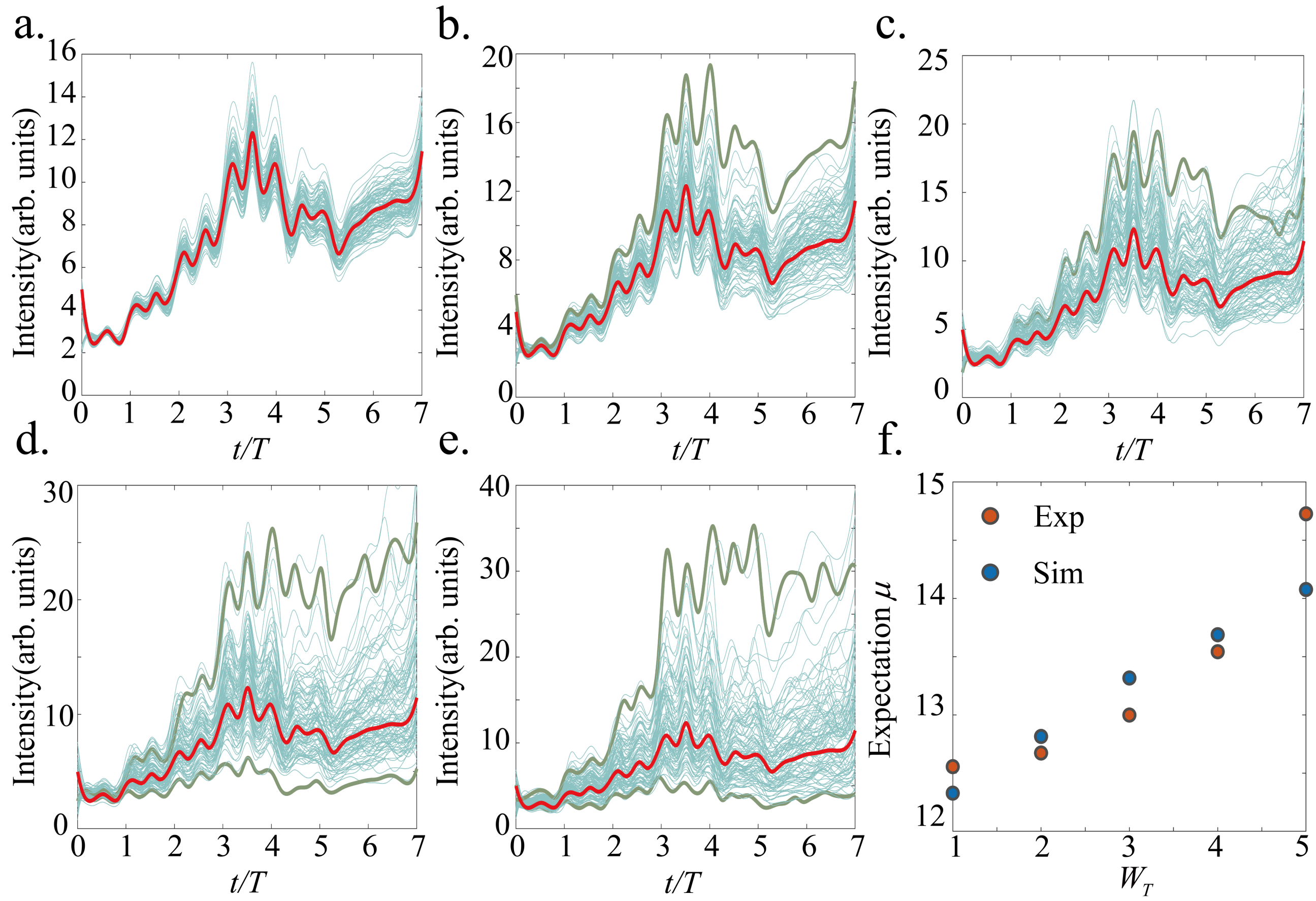

Extended Data Figure 2. Experimental dynamical responses under FRTN. **a-e**, Dynamic responses from individual experimental realizations corresponding to noise strengths $W_T = 1$ (**a**), $W_T = 2$ (**b**), $W_T = 3$ (**c**), $W_T = 4$ (**d**), $W_T = 5$ (**e**). The red line designates the deterministic response of the noiseless system. Driven by temporal noise, the responses fluctuate randomly around the noiseless baseline. **f**, Comparison of the intensity expectations between the experimental measurements and numerical simulations as a function of $W_T$.